# Enhancement of Mechanical Properties of Graphene Oxide Fibers via Liquid Crystalline Phase Formation and Flake Size Optimization

M. Zhezhu[1], G. Baghdasaryan[1], G. Gevorgyan[2], S. Gyozalyan[1], Y. Grigoryan[1], K. Margaryan[2], H. Gharagulyan[1, 2*]

[1] *Liquid Crystalline Nanosystems Research Group, A.B. Nalbandyan Institute of Chemical Physics NAS RA, Yerevan 0014, Armenia*
[2]*Institute of Physics, Yerevan State University, Yerevan 0025, Armenia*

[*]*Author to whom correspondence should be addressed:* herminegharagulyan@ysu.am

**Abstract**

**Graphene oxide (GO) fibers are promising materials for lightweight, high-strength applications due to their unique structural tunability and mechanical performance. However, the properties of GO fibers strongly depend on the ordering of GO flakes during the assembly process. In this work, we demonstrate that GO fibers spun from a liquid crystalline (LC) GO dispersion exhibit significantly enhanced mechanical properties compared to those produced from non-LC GO dispersions. The improved tensile strength is attributed to the larger GO flake size and highly ordered alignment achieved in the LC phase. The LC-derived fibers demonstrated a Young's modulus of 12.3 GPa, a tensile strength of 146.8 MPa, and an elongation at break of 2.5%. These findings emphasize the critical role of flake size and LC ordering in enhancing the performance of GO-based fibers and suggest a straightforward pathway toward scalable fabrication of strong yet flexible carbon-based materials.**

## 1. Introduction

Graphene oxide (GO) fibers are promising lightweight materials with high strength and flexibility [1]. Yet, their performance is largely determined by flake alignment during spinning and the characteristics of the polymer matrix [2]. Depending on concentration and flake aspect ratio, GO can form liquid crystalline (LC) phases [3], which promote highly aligned and densely packed fiber formation for applications in robotic actuators [4], textile electronics [5], and beyond. Nevertheless, systematic comparisons between LC and non-LC GO fibers remain limited, as do studies on fibers produced from GO dispersions with varying flake sizes.

In this study, GO was synthesized *via* electrochemical exfoliation and subsequently fractionated by centrifugation to obtain dispersions with different flake sizes. Among the separated fractions, one displayed clear evidence of liquid crystalline phase formation. Fibers were fabricated through wet-spinning using an alginate matrix. The resulting fibers were comprehensively characterized in comparison with the initial GO dispersion. Detailed analyses of the microstructure of fibers from both LC and non-LC dispersions were performed, and their mechanical properties were thoroughly evaluated. Although prior works have demonstrated that LC ordering in GO dispersions enables macroscopic alignment and improved mechanical properties compared to isotropic spinning [6], a major unresolved bottleneck is insufficient stress transfer across flake interfaces due to limited long-range LC domain size, residual misalignment at macro-scales. In particular, uncontrolled flake size distribution leads to high defect density and poor percolation of load-bearing/transferring pathways. This study systematically investigates the effect of controlled flake size distribution and LC domain formation, alignment quality, and macroscopic mechanical properties.

## 2. Materials and methods

GO was synthesized *via* electrochemical exfoliation [7, 8]. The initial dispersion of GO had a concentration of 0.01 mg/mL. Samples GO-1, GO-2, GO-3, and GO-4 were obtained from sequential fractionation of the initial GO dispersion by centrifugation at 8000 rpm for 1, 5, 10, and 15 minutes, respectively. Thus, the sequential fractionation is expected to result in a gradual decrease in the GO flake sizes, with GO-1 exhibiting the largest size and GO-4 the smallest. Each 5 mL portion of the GO dispersion was mixed with a few drops of ammonia solution and 5 mL of NaAlg solution. The same centrifugation protocol was applied to the liquid crystalline graphene oxide (LCGO) sample, specifically at a rotational speed of 5000 rpm for 5 minutes. The critical concentration required for LC formation was 2.5mg/mL. Lateral size of GO flakes which gave LC phase was about 10 μm, and the thickness ≈2 nm [3]. The fibers were formed by wet-spinning into a 3% $CaCl_2/H_2O$-ethanol (3:1) coagulation bath. Then they were immediately removed from the bath, rinsed, and dried at 40°C for 10 minutes under vacuum. The fibers were numbered GOF-1, GOF-2, GOF-3, and GOF-4, according to the corresponding GO fraction used for their preparation, while the fiber from the LCGO sample was labeled as $GO_{LC}F$.

The liquid crystalline phase of GO was examined by polarized optical microscope (POM) (MP920, BW Optics). Optical properties of GO dispersions were studied by UV–Vis spectroscopy (Cary 60, Agilent), while chemical bonding was analyzed by FTIR-ATR (Spectrum Two, PerkinElmer) and Raman spectroscopy (LabRAM HR Evolution, HORIBA). Thermal behavior was evaluated using TGA/DSC (TGA/DSC 3+, Mettler Toledo) under an argon atmosphere. Morphology was characterized by SEM (Prisma E, Thermo Fisher), particle size by the DLS method (Litesizer 500, Anton Paar), and mechanical properties by a tensile testing system (HCS350G-TNS, Instec).

## 3. Results and discussion

Fabrication stages are shown in Fig. 1a. Before $GO_{LC}F$ spinning, POM confirmed the LC nematic phase of GO (Fig. 1b), supported also by ellipsometry in [8]. A detailed LC domain analysis was performed based on POM images, indicating that, while the overall distribution across the cell is almost homogeneous, the domains remain partially localized. The clear preferable direction of flakes is observed (Fig. S1).

SEM analysis revealed fiber diameters of 80-130 μm. The typical morphologies (Fig. 1(c-e)) show GOF with a wrinkled surface, while fibers from the smallest flakes display a non-porous structure. The $GO_{LC}F$ exhibits a wavy texture with well-aligned fine filaments forming the fiber.

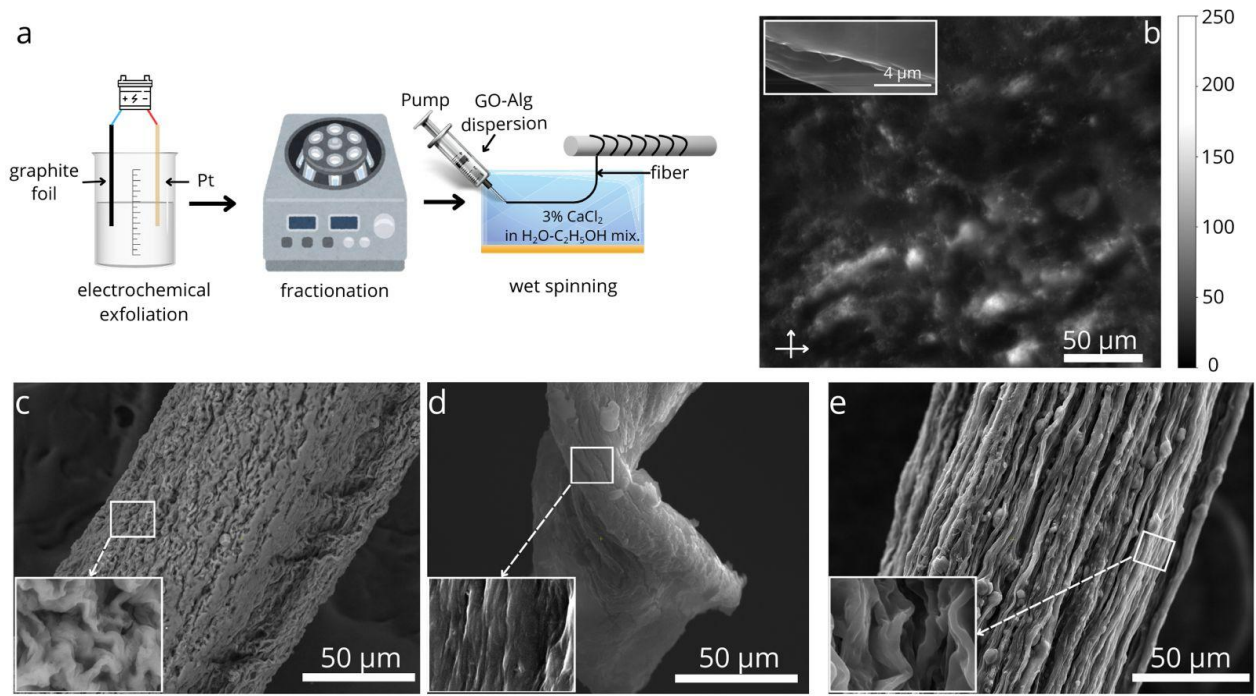

Fig. 1. (a) Schematic of the GOF fabrication process; (b) LC phase of GO dispersion with the color bar indicating birefringence intensity (inset: SEM image of LCGO flakes); SEM images of (c) GOF-1, (d) GOF-4, and (e) $GO_{LC}F$.

A detailed characterization of GO dispersion and spun fiber is provided in the Supplementary Materials (Figs. S2 and S3, with peak assignments listed in Table S1).

The influence of LCGO on fiber formation is shown in Fig. 2(a–d). The $GO_{LC}F$ has a ~130 μm cross-section and a well-organized structure of aligned 0.4–1.7 μm filaments, characteristic of LC-directed assembly. In contrast, GOF-1 (~110 μm cross-section) displays a rough, more disordered packing. These differences demonstrate the ordering effect of the LC phase.

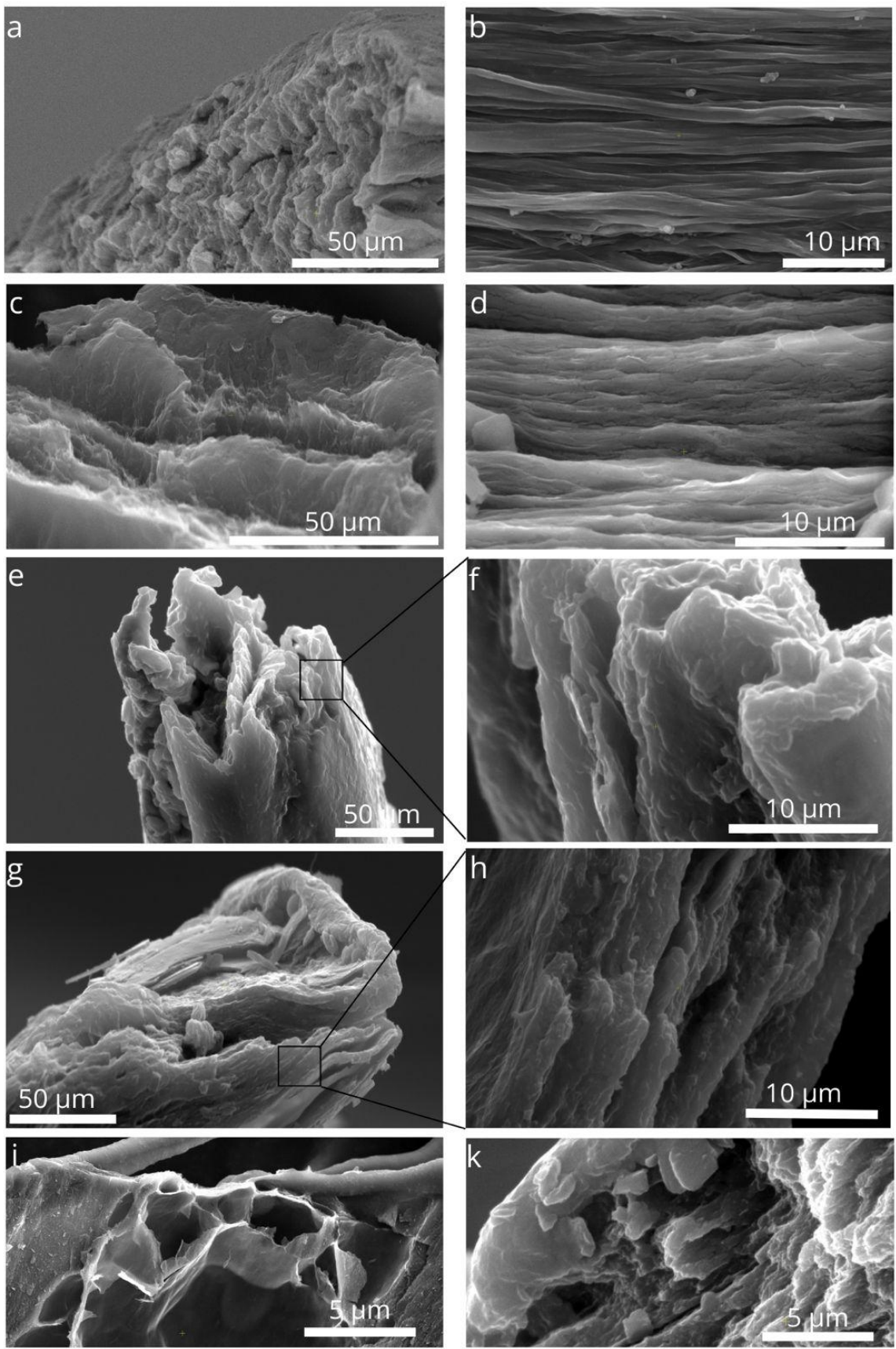

Fig. 2. SEM images of (a) cross-sections of GO$_{LC}$F and (b) GOF; filament alignment within the fibers for (c) GO$_{LC}$F and (d) GOF; cross-sections of fractured (e) GO$_{LC}$F and (g) GOF; interlayer sliding in fractured (f) GO$_{LC}$F and (h) GOF; and porous structures in the cross-sections of fractured (i) GO$_{LC}$F and (k) GOF.

As evidenced from post-fracture SEM images (Fig.2 (e-k)), larger LC domains and higher macroscopic alignment minimize misalignment defects, reduce the density of weak inter-flake junctions, and promote more efficient load distribution which, in turn, affect delayed crack initiation before failure [9].

The mechanical testing of GOF with different flake sizes (Fig. 3) shows that $GO_{LC}F$ outperforms non-LC GOFs, exhibiting higher stress at break and greater Young's modulus due to larger flakes and improved alignment.

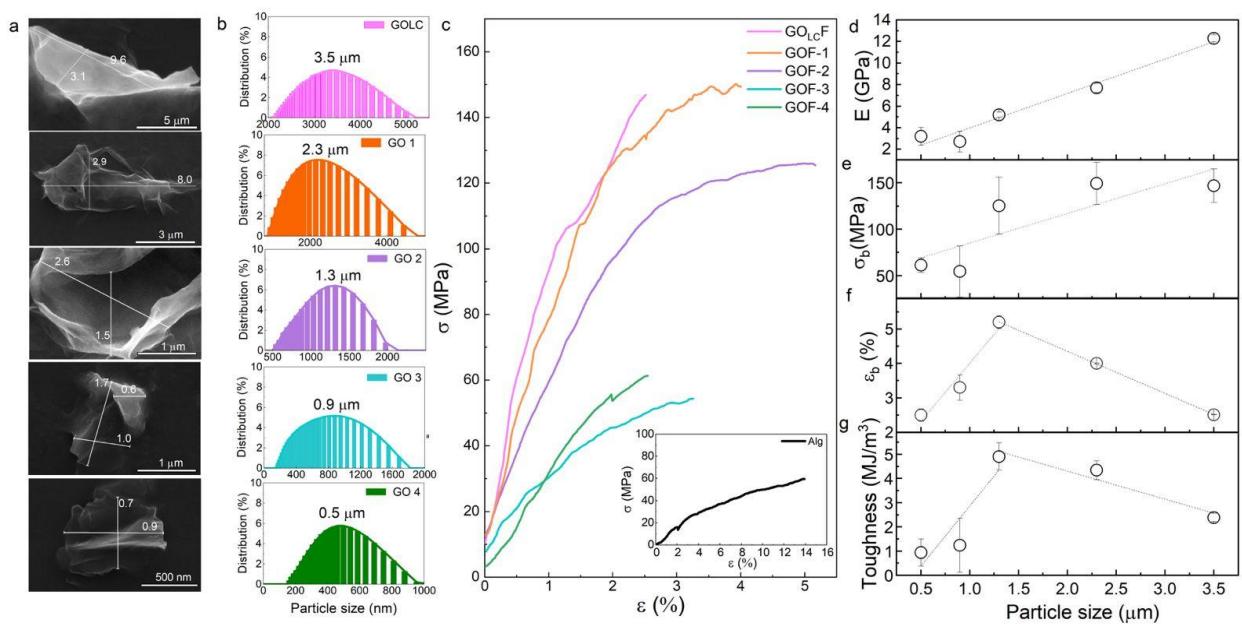

Fig. 3. (a) SEM images of typical GO flakes in the initial GO dispersions before fiber fabrication; (b) particle size distribution measured by DLS. Mechanical properties of fibers (inset: reference alginate fiber): (c) stress–strain curves; (d) Young's modulus ($E$); (e) stress at break ($\sigma_\beta$); (f) elongation at break ($\varepsilon_\beta$); and (g) toughness. The tensile rate was set at 3 μm/s, and the measurement error was ±1 %. For each set of experimental conditions, three replicates were conducted to ensure the reliability of the results.

The tensile strength increases approximately linearly with the flake size, indicating more efficient stress transfer between well-aligned, larger GO sheets. In contrast, $\varepsilon_b$ and toughness vary non-linearly, suggesting competing effects from alignment and interfacial interactions. It is also noteworthy that fibers composed of particles >0.9 μm demonstrate greater toughness, primarily due to the alginate's plasticizing and cross-linking effects, while very fine particles (<0.9 μm), create more interfaces and stress concentration points within the fiber, reducing $\varepsilon_b$ and toughness. Furthermore, particle sizes between 1.3 μm and 2.3 μm provide an optimal balance and yield the highest mechanical performance.

Long-term stability test (Fig. 4(a–d)) shows that $GO_{LC}F$ maintains its structure, with minimal nanospherical alginate agglomeration. Non-LC GO fibers, however, accumulate surface agglomerates and gradually flatten, indicating poorer resistance to chemical aging.

After one year, the mechanical properties of the fibers changed moderately (Fig. 4(e-f)). The Young's modulus of $GO_{LC}F$ decreased significantly, while all fibers retained $E$ above 2.2 GPa. Trends in $E$ and $\sigma_\beta$ were preserved, with $\sigma_\beta$ decreasing from ~147 MPa ($GO_{LC}F$) and 149 MPa (GOF-1) to 132 MPa and 105 MPa, respectively. Changes in $\varepsilon_\beta$ and toughness were minor, with maximum values still observed for GOF-3 (particle size of 1.3 μm).

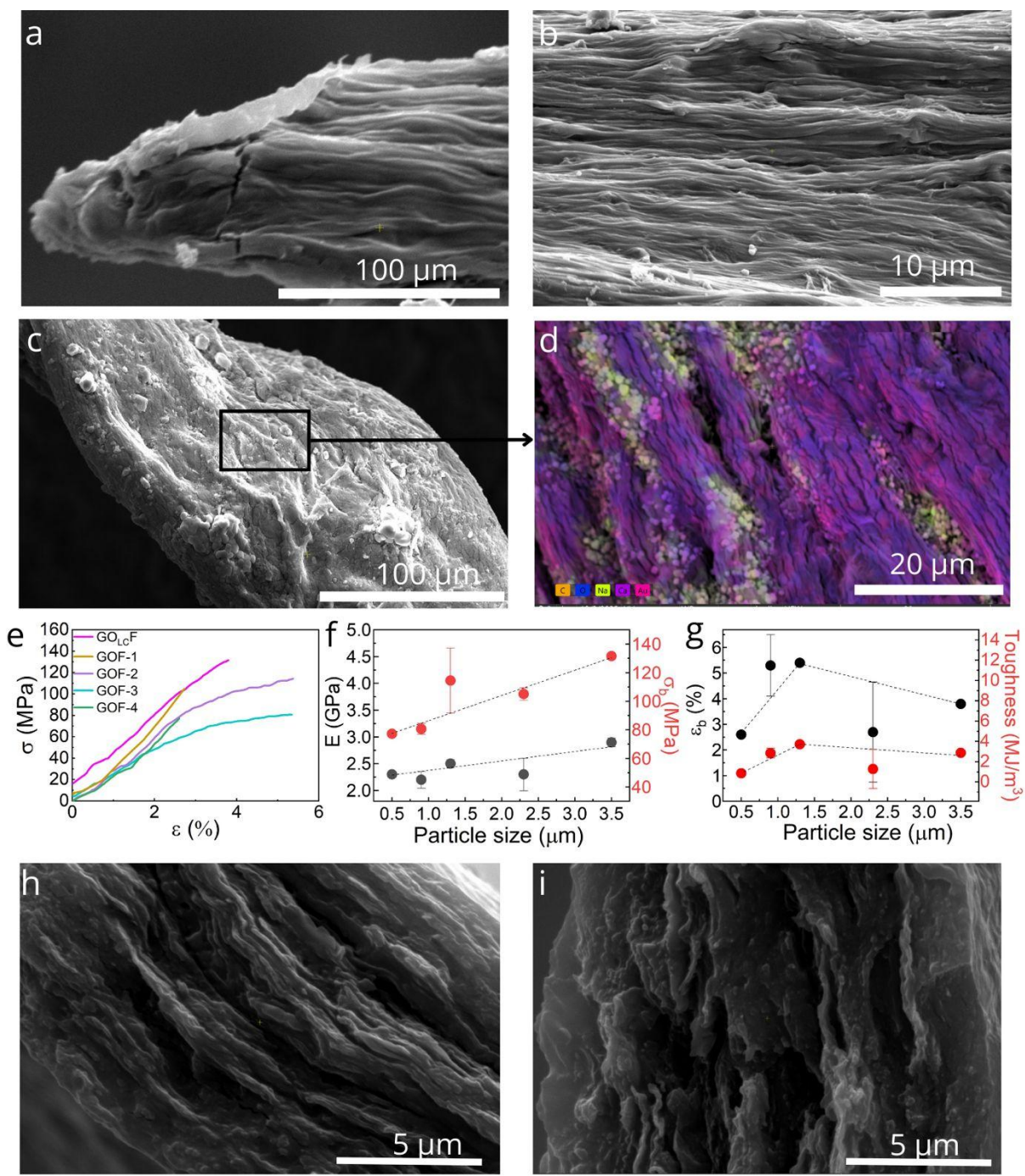

Fig. 4. (a) The GO$_{LC}$F and (b) its filament ordering after 6 months. (c) GOF after 6 months with (d) EDS surface mapping. Elemental distribution is shown as follows: carbon – orange, oxygen – blue, sodium – yellow-green, calcium – purple, and gold – pink (the latter corresponds to the gold coating applied before SEM analysis). Mechanical properties from repeated measurements 1 year after fiber fabrication: (e) stress–strain curve, (f) $E$ and $\sigma_b$, (g) $\varepsilon_b$ and toughness, interlayer sliding after 1 year for (h) GO$_{LC}$F and (i) GOF.

Meanwhile, the fiber surfaces appeared slightly shrunken (Fig. 4(h, i)), likely as a result of moisture loss, while the filament arrangement remained largely intact. The average fiber diameter showed only a modest reduction, decreasing by no more than ~15% for all samples.

Literature data for wet-spun GO-based fibers are presented in Table S2.

## 4. Conclusion

This work highlights the critical influence of GO flake size and organization on the mechanical properties of GO fibers. Fibers produced from LCGO dispersions exhibit noticeably higher stiffness and tensile strength than those from non-LC dispersions, primarily due to the larger flake size and improved structural alignment within the LC phase. The LC-derived fibers achieved a Young's modulus of 12.3 GPa, tensile strength of 146.8 MPa, and elongation at break of 2.5%. However, the key advantage of our fibers lies in the ability to tune and optimize their mechanical properties, specifically flexibility and tensile strength, by controlling the flake size. Moreover, in the case of $GO_{LC}F$, these parameters reach an intrinsic equilibrium, resulting in a balanced combination of strength and toughness. These findings demonstrate that controlling flake size and LC ordering provides a simple and scalable strategy for fabricating strong, lightweight, and flexible GO-based fibers for potential use in advanced mechanical and functional applications. The long-term stability and aging behavior of the fibers were investigated after 6 and 12 months of storage under ambient conditions.

# Supplementary Materials

for

## Enhancement of Mechanical Properties of Graphene Oxide Fibers via Liquid Crystalline Phase Formation and Flake Size Optimization

M. Zhezhu[1], G. Baghdasaryan[1], G. Gevorgyan[2], S. Gyozalyan[1], Y. Grigoryan[1], K. Margaryan[2], H. Gharagulyan[1, 2*]

[1] *Liquid Crystalline Nanosystems Research Group, A.B. Nalbandyan Institute of Chemical Physics NAS RA, Yerevan 0014, Armenia*
[2] *Institute of Physics, Yerevan State University, Yerevan 0025, Armenia*

[*]*Author to whom correspondence should be addressed:* herminegharagulyan@ysu.am

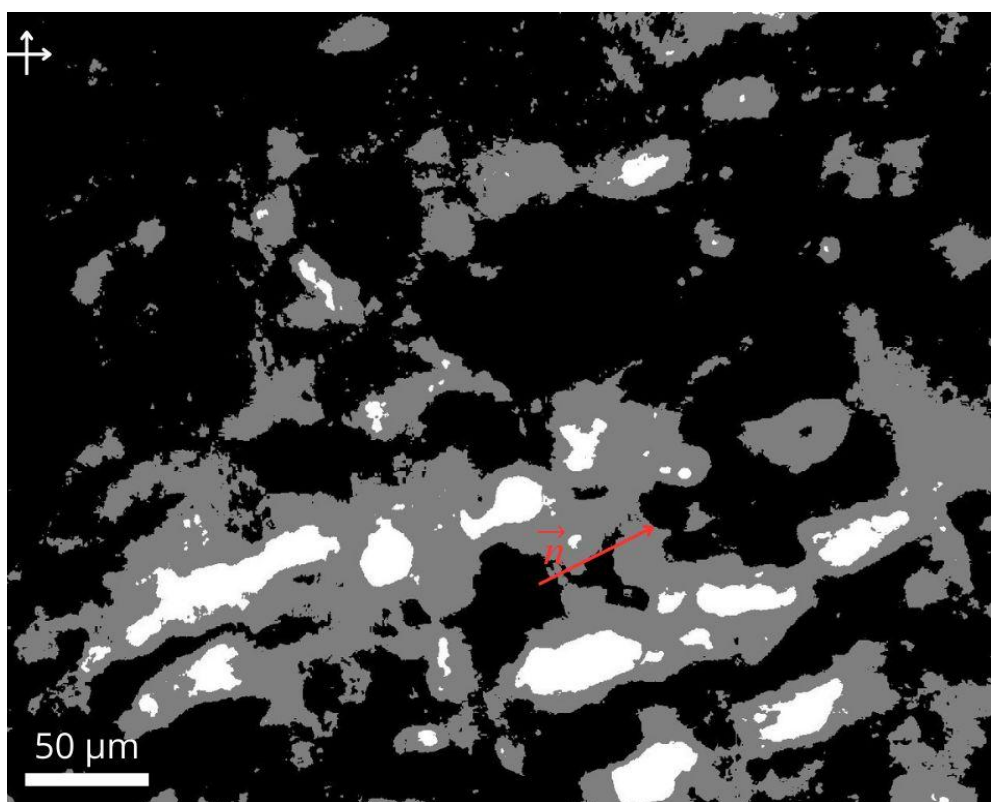

Fig. S1. LC domain analysis of the LCGO POM image: K-means clustering (k=3) segmented the image into dark (66.75%), gray (28.53%), and bright regions (4.72%). The higher percentage observed for the dark region is due to fewer clusters; for more precise estimation, a larger number of clusters should be considered. The director is indicated by an arrow.

**Characterization of GO dispersion and spun GOF-1**

Analysis of GOF relative to the GO dispersion is shown in Fig. S2. UV–Vis spectra of initial GO exhibit bands at 238.2 nm (π→π* transition of C=C) and 296.5 nm (n→π* transition of C=O groups) [S1]. In the GO-alginate (GO-Alg) dispersion, peaks at 237.7 nm and 291.1 nm indicate a slight red shift, suggesting interactions between GO and the alginate matrix, most likely *via* hydrogen bonding or partial modification of GO's surface groups.

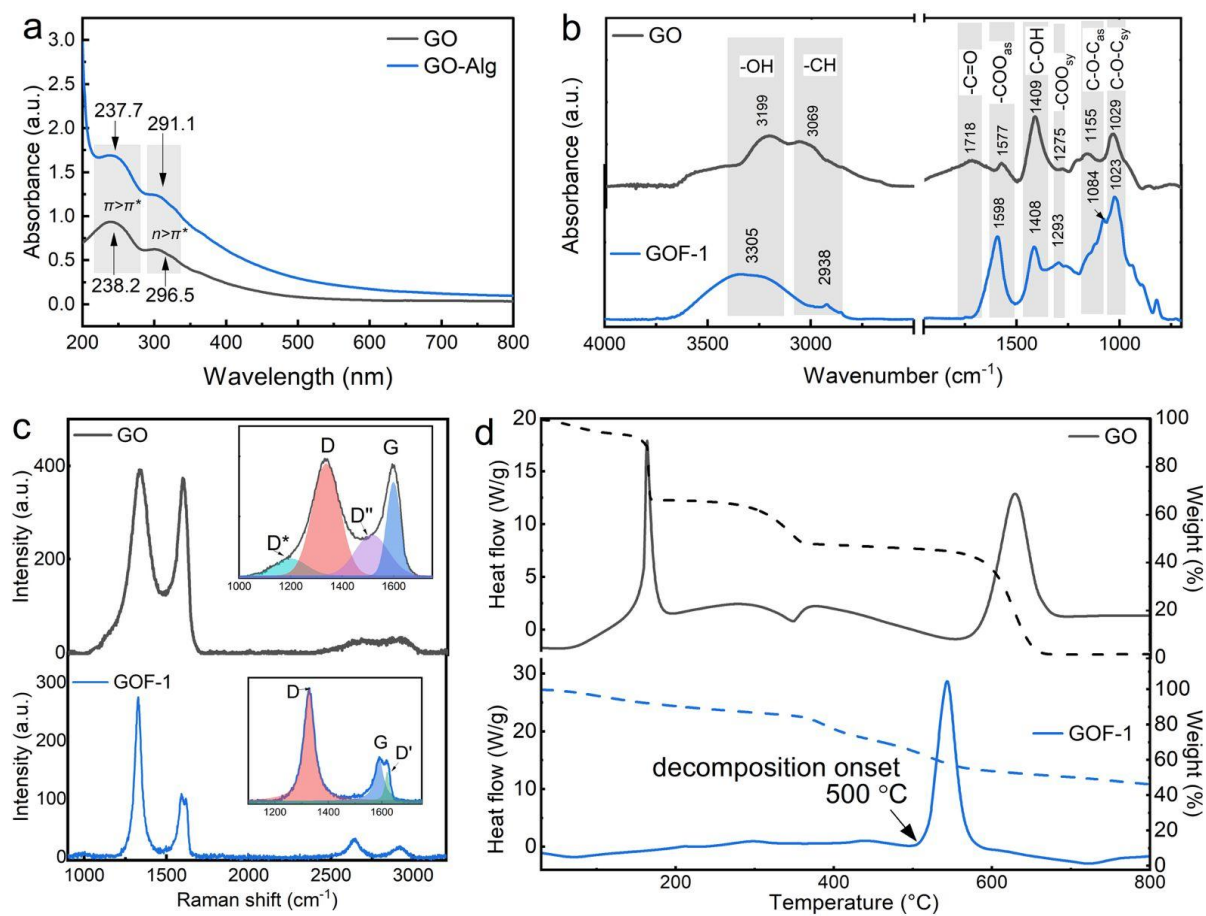

Fig. S2. (a) UV–Vis absorption spectra of GO and GO-Alg in aqueous solution. Comparative characterization of the initial GO powder and the GOF-1 obtained from it by (b) FTIR–ATR, (c) Raman spectroscopy, and (d) DSC–TGA analysis.

FTIR-ATR spectra of GO, NaAlg (Fig. S3), and GOF (peak assignments in Table S1) reveal several changes confirming these interactions. In GOF, the broad O–H band shifts (~3305 cm$^{-1}$) and widens, the carbonyl peak at 1718 cm$^{-1}$ disappears with a stronger -COO$^-$ band at 1598 cm$^{-1}$, suggesting hydrogen/ionic bonding with NaAlg and possible partial GO reduction. Shifts in the 1200–800 cm$^{-1}$ region indicate overlapping C–O–C and C–O vibrations originating from both GO and NaAlg. Overall, the GOF spectrum combines features of GO and NaAlg with modifications indicative of hydrogen bond formation during the fiber fabrication.

Raman spectra of GO-based materials typically show the D, G, 2D, and D+G bands. Fitting in the 1000–1800 cm$^{-1}$ region of pure GO also revealed D* and D″ features, linked to edge-related sp$^2$–sp$^3$ disorder and amorphous or interstitial defect [S2, S3]. In the GOF, a small D′ peak appears, characteristic of 5–8–5 defects in wrinkled few-layer structures [S2]. Table S1 lists the peak positions and assignments. The fiber's higher $I_D/I_G$ ratio (1.75 vs 1.05 for GO) indicates an increased defect density, likely from defect formation on the graphitic flakes and new sp$^2$ domains development.

The thermal analysis shows that GOF exhibits higher thermal stability compared to individual GO, remaining stable up to 500 °C (Fig. S2 and Fig. S3). The delayed onset of decomposition and reduced total weight loss suggest strong interfacial interactions and improved structural integrity within the fiber. The detailed thermal analysis is provided below.

Table S1. FTIR-ATR, Raman analysis of GO, NaAlg, and GO–Alg showing characteristic peak wavenumbers and their corresponding assignments.

| FTIR-ATR | | | | |
|---|---|---|---|---|
| GO | NaAlg | GO-Alg fiber | Functional group | References |
| Wavenumber, cm$^{-1}$ | | | | |
| 3199 | 3294 | 3305 | -OH | [S2-S8] |
| 3069 | - | - | C-H | |
|  | 2905 | 2938 | | |
| 1718 | - | - | C=O | |
| 1577 | 1594 | 1598 | COO$^-$ as st | |
| 1409 | 1400 | 1408 | C-OH | |
| 1275 | 1294 | 1293 | COO$^-$ sy st | |
| 1155 | 1071 | 1084 | C-O-C as st in GO; C-O-C st in NaAlg | |
| 1029 | 1025 | 1023 | C-O-C sy st in GO; C-O st in NaAlg | |
| **Raman spectroscopy** | | | | |
|  |  | GO | GO-Alg | Assignment |
| Peak position, cm$^{-1}$ | | 1200.5 | - | D* |
| | | 1338.0 | 1327.6 | D |
| | | 1514.7 | - | D'' |
| | | 1598.8 | 1590.4 | G |
| | | - | 1619.7 | D' |

**Thermal analysis**

The thermal decomposition of NaAlg involves a series of exothermic and endothermic processes. The first thermal process appears around 98 °C as an endothermic peak, accompanied by a 13% weight loss, corresponding to the evaporation of physically adsorbed and bound water [S8, S9]. This is followed by two gradual exothermic decomposition steps at 254 °C and 368 °C, resulting in a combined weight loss of 52%. These processes are attributed to the cleavage of glycosidic linkages, dehydration, decarboxylation, and decarbonylation of the alginate backbone, followed by the decomposition of intermediate carbonaceous products [S9]. The thermal process observed between 562 °C and 654 °C results in an additional 15% weight loss. This process is likely associated with further carbonization and partial combustion of the residual organic matter, along with the thermal degradation of sodium-containing species to form carbonized structures, $Na_2CO_3$, $Na_2O$, NaOH [S10]. The total mass loss during the decomposition process was approximately 83%.

In contrast to the individual components, the GOF demonstrates a distinct thermal degradation behavior. The first endothermic event occurs at 72 °C and corresponds to the loss of surface moisture. This temperature is higher than that observed for pure GO, indicating enhanced water retention, likely due to the hydrophilic alginate matrix.

Importantly, the onset of thermal decomposition in the fiber begins at 211 °C, higher than the 165 °C observed for pure GO. This shift may be partially attributed to interfacial interactions between GO and the alginate matrix. During thermal annealing, chemical interactions between the two components may result in the formation of new bonds, which in turn prolong the degradation process. Additionally, the geometry of the fiber may play a role, as the thermal reduction of bulk GO is known to be geometry-dependent, influencing parameters such as peak decomposition temperature, reaction rate, and the chemical composition of evolved gases [S11].

Further, the degradation is accompanied by several low thermal events at 299 °C and 440 °C, associated with the destructurization of the alginate matrix. A strong exothermic peak at 543 °C, the most intense feature in the GOF profile, is likely associated with the main decomposition process. Finally, an endothermic peak at 721 °C leads to a further 6% mass loss, possibly due to the degradation of thermally stable inorganic species.

Overall, the total weight loss of the GOF is 79%, which is lower than that observed for both pure GO and NaAlg. This reduction may be due to the formation of residual carbon, $CaCO_3$, and CaO as a result of decarboxylation processes [S12].

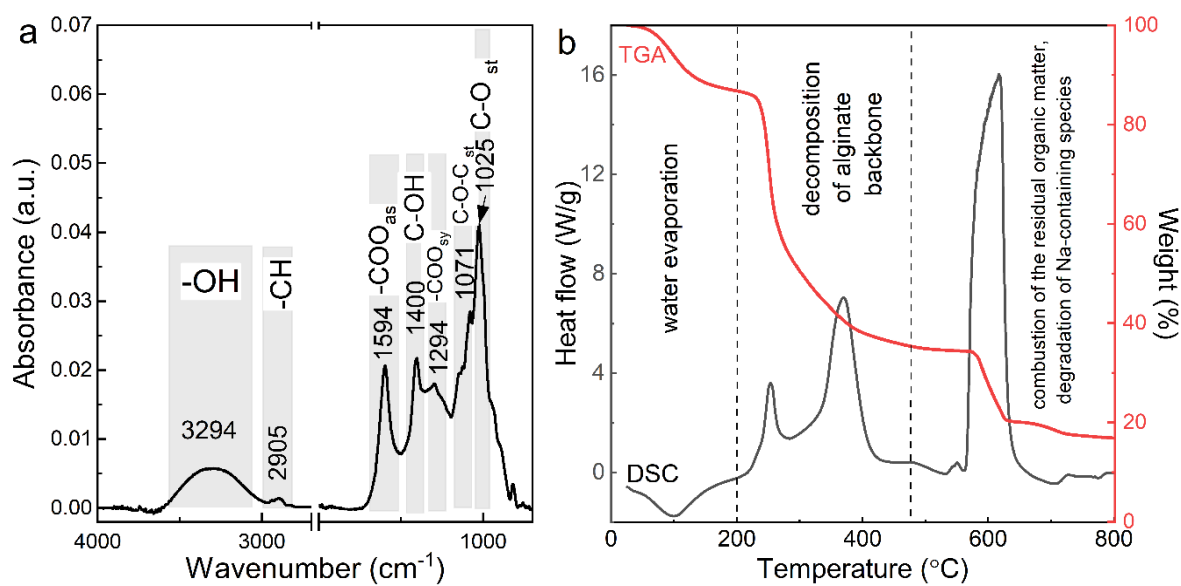

Fig. S3. NaAlg characterization by (a) FTIR–ATR and (b) DSC–TGA analysis.

Table S2. Summary of literature data for GO and GO-based fibers prepared by wet spinning with different coagulation baths.

| Fiber | LC dispersion | Coagulation bath | Mechanical properties | | | | Reference |
|---|---|---|---|---|---|---|---|
| | | | $E$, GPa | $\varepsilon_b$, % | $\sigma_b$, MPa | *Toghness*, MJ/m$^3$ | |
| GO/PEDOT:PSS 8:2 wt/wt | yes | 70 wt% acetic acid solution with calcium acetate | 0.3 | ~11 | >20 | – | [S9] |
| PEG-GO/PEDOT:PSS | yes | | 0.04 | 18 | >10 | – | |

| GO | yes | CaCl$_2$ | ~5.4 | ~6.1 | ~62.9 | ~2.9 | [S13] |
|---|---|---|---|---|---|---|---|
| GO | yes | KOH | ~2.9 | ~3.6 | ~41.2 | ~1.0 | |
| GO | yes | CaCl$_2$ up to 10% wt. | ~6.8 | ~4.9 | ~108.9 | 3.7 | |
| GO | yes | CaCl$_2$ | 6.3 | 6.8 | 364.4 | – | [S12] |
| GO | yes | CuSO4 | 6.4 | 5.9 | 274.3 | – | |
| GO | yes | KOH | 3.2 | 7.5 | 184.6 | – | |
| GO | yes | CaCl$_2$ | 20.1 | ~3.2 | 412 | 4.8 J/g$^{-1}$ | [S14] |
| GO | yes | chitozan | 22.6 | ~3.5 | 442 | 4.8 J/g$^{-1}$ | |
| 4 wt.% GO/NaAlg fibers | no | CaCl$_2$ | 4.3 | 14 | 620 | – | [S15] |
| GO$_{LC}$F | yes | 3% CaCl$_2$/H$_2$O-ethanol (3:1) | 12.3 | 2.5 | 146.8 | 2.39 | this study |
| GOF-1 | no | | 7.7 | 4 | 149.3 | 4.34 | |

The comparative analysis of GO fibers demonstrates clear distinctions between fibers obtained from LC dispersions and those spun from non-LC dispersions. Literature data indicate that LC GO fibers exhibit various mechanical properties, with Young's modulus spanning 0.04–22.6 GPa, elongation at break ranging from 3.2% to 18%, stress at break from 10 to 442 MPa, and toughness between 1 and 3.7 MJ/m3. In contrast, a non-LC GO fiber, such as 4 wt% GO/NaAlg, shows E = 4.3 GPa, ε_(b ) = 14%, and σ_b = 620 MPa. These comparisons confirm that LC GO fibers generally achieve higher Young's modulus, whereas non-LC fibers display greater elongation and stress at break. The results of this study reveal that GOLCF exhibits a higher Young's modulus than the majority of LC GO fibers reported in the literature, while non-LC GOF also shows relatively elevated Young's modulus values. Overall, the mechanical properties measured in this work are in good agreement with previously reported data, with minor deviations likely arising from variations in coagulation bath composition, reactant concentrations, and other procedure parameters. Importantly, these findings emphasize the predominant role of the LC phase in producing fibers with enhanced stiffness.